\documentclass{article}

\pdfoutput=1

\usepackage{bm,latexsym,amsmath,amssymb,amsfonts,fancyhdr,color,graphicx,multirow}
\usepackage[a4paper,bottom=2cm,top=2.5cm,head=5mm,width=18cm,dvipdfm]{geometry}
\usepackage[usenames,dvipsnames,svgnames,table]{xcolor}
\usepackage[colorlinks=true,
            linkcolor=blue,
            urlcolor=blue,
            citecolor=green,          
						bookmarks=true,
						bookmarksnumbered=true,
						breaklinks=true,
						pdfpagemode=Fullscreen,
						pdfstartview=FitBH]{hyperref}
\usepackage[dotinlabels]{titletoc}
\usepackage{titlesec}
\usepackage{authblk}

\numberwithin{equation}{section}
\allowdisplaybreaks[4]

\titlelabel{\thetitle.\quad \hspace{-0.8em}}
\titlecontents{section}
              [1.5em]
              {\vspace{4mm} \large \bf}
              {\contentslabel{1em}}
              {\hspace*{-1em}}
              {\titlerule*[.5pc]{.}\contentspage}
\titlecontents{subsection}
              [3.5em]
              {\vspace{2mm}}
              {\contentslabel{1.8em}}
              {\hspace*{.3em}}
              {\titlerule*[.5pc]{.}\contentspage}
\titlecontents{subsubsection}
              [5.5em]
              {\vspace{2mm}}
              {\contentslabel{2.5em}}
              {\hspace*{.3em}}
              {\titlerule*[.5pc]{.}\contentspage}

\newcommand{\titledef}{Constraining Photon Portal Dark Matter with TEXONO and COHERENT Data} 

\hypersetup{ pdfauthor = {Shao-Feng Ge},
	     pdftitle = {\titledef}, 
	     pdfsubject = {}, 
             pdfkeywords = {}, 
	     pdfcreator = {LaTeX with hyperref package}}

\definecolor{gesfpurple}{rgb}{0.47,0.19,0.42}

\definecolor{gesflanse}{rgb}{0.00,0.50,0.50}

\definecolor{gesfblue}{rgb}{0.08,0.42,0.76}

\definecolor{gesfred}{rgb}{1,0,0}

\definecolor{gesfwhite}{rgb}{1,1,1}

\definecolor{gesfblack}{rgb}{0,0,0}

\newcommand{\geqn}[1]{\hypersetup{linkcolor=blue}(\ref{#1})\hypersetup{linkcolor=blue}}
\newcommand{\gfig}[1]{{\hypersetup{linkcolor=violet}Fig.~\ref{#1}\hypersetup{linkcolor=blue}}}

\renewcommand*{\tablename}{Table}
\renewcommand*{\figurename}{Fig.}

\usepackage{mathrsfs,amssymb}  
\usepackage{mathtools}

\newcommand{\be}{\begin{equation}}
\newcommand{\ee}{\end{equation}}
\usepackage{slashed}

\graphicspath{{figs/}}

\begin{document}
\fontsize{12pt}{14pt}\selectfont

\title{
       \textbf{\huge \titledef}} 
\author[1]{{\large Shao-Feng Ge} \footnote{gesf02@gmail.com}}
\affil[1]{\small Kavli IPMU (WPI), UTIAS, The University of Tokyo, Kashiwa, Chiba 277-8583, Japan}
\affil[1]{\small Department of Physics, University of California, Berkeley, CA 94720, USA}
\author[2]{{\large Ian M. Shoemaker} \footnote{ian.shoemaker@usd.edu}}
\affil[2]{\small Department of Physics, University of South Dakota, Vermillion, SD 57069, USA}
\date{\today}

\maketitle

\vspace{-95mm}
\hfill IPMU17-0149
\vspace{85mm}
\begin{abstract}
Dark Matter may reside in sector without Standard Model (SM) gauge interactions.  One way in which such a dark sector can still impact SM particles through non-gravitational interactions is via the ``photon portal'' in which a dark photon kinetically mixes with the ordinary SM photon. We study the implications of this setup for electron recoil events at TEXONO reactor and nuclear recoil events at the COHERENT experiment. We find that the recent COHERENT data rules out previously allowed regions of parameter space favored by the thermal relic hypothesis for the DM abundance. When mapped onto the DM-electron cross section, we find that COHERENT provides the leading direct constraints for DM masses $<$ 30 MeV.

\end{abstract}

\section{Introduction}

The majority of the matter in our Universe is non-luminous and non-baryonic. To date, all evidence of this Dark Matter (DM) has been gravitational in nature. Many models of new physics predict DM candidates with additional interactions beyond gravity. Indeed, one of the most studied frameworks for explaining the DM abundance posits that some new non-gravitational interactions brought DM into thermal equilibrium in the early Universe through 2-to-2 annihilation processes. Eventually, Hubble expansion dilutes the DM density so much that the annihilation rate plummets and the abundance of DM ``freezes-out.'' The relic abundance of such thermal DM can easily be in line with observations for cross sections $\langle \sigma v\rangle_{{\rm ann}} \simeq 6 \times 10^{-26}~{\rm cm}^{3}~{\rm s}^{-1}$, with a weak dependence on the mass of the DM. This is the classic hypothesis for DM as a Weakly Interacting Massive Particle (WIMP), which appears in many extensions of the SM.  Being charged under electroweak interactions, however, the WIMP hypothesis leads to a number of predictions, which to date have only been tightly constrained. 

A natural question is then: can the thermal relic hypothesis for DM survive beyond WIMPs? Indeed, a simple and theoretically motivated scenario is one in which DM and perhaps a whole array of new particles--a ``hidden'' or ``dark'' sector--shares no gauge interactions with the SM. In lieu of gauge interactions, the visible and hidden sectors may communicate through gauge invariant combinations of the fields in the two sectors. At the renormalizable level there are a surprisingly small number of options for such ``portals''
\begin{equation}
    \mathscr{L}_{{\rm portal}} =
    \begin{cases*}
      \epsilon F_{\mu \nu} F_{h}'^{\mu \nu}  & ({\rm photon~portal})\\
      h |H^{2}| |H_{h}^{2}|       &  ({\rm Higgs~portal})\\
      y (LH) N      & ({\rm neutrino~portal}),
    \end{cases*}
  \end{equation}
where $F'_{\mu \nu}$, $H_{h}$, and $N$ are respectively hidden sector field strengths, Higgses, and fermions. Typically the impact of each of these portals is separately treated, as each one leads to distinct search strategies. 

In this paper we study the impact of the photon portal for light DM, in which the SM photon kinetically mixes with a $U(1)$ dark photon \cite{Holdom:1985ag}. The implications of $\gamma-\gamma'$ kinetic mixing for DM has been widely studied~\cite{Foot:2004pa,Feldman:2006wd,ArkaniHamed:2008qn,Pospelov:2008jd,Bjorken:2009mm,Davoudiasl:2013jma,Campos:2017dgc}. At the phenomenological level, the photon portal gives rise to two main classes of probes: (1) direct detection signals from DM-proton or DM-electron scattering, and (2) the production of DM at accelerators and colliders. Given the strong direct detection constraints, we will focus on the sub-GeV regime for DM.  Notice that the strength of the direct detection constraints for $>$ GeV DM masses is partly thanks to the coherent enhancement of the DM scattering on the nucleus. 

In light of the recent discovery of coherent neutrino-nucleus scattering~\cite{Akimov:2017ade} by the COHERENT collaboration, we ask what the COHERENT data brings to bear on photon portal models of light DM. The possibility of producing and detecting light DM at coherent neutrino-nucleus experiments was studied in~\cite{deNiverville:2015mwa}. We additionally study the ability of reactor neutrino experiments like TEXONO to constrain light DM from their electron recoil events. The mass reach of TEXONO extends to $\sim$ MeV masses, while COHERENT's stopped pion source can access DM masses out to $\sim$ 65 MeV. 

The remainder of this paper is organized as follows. In Sec.~\ref{model} we introduce the model of study with a kinetically mixed dark photon interacting with pairs of DM particles. In Sec.~\ref{tex} we examine the sensitivity at TEXONO to dark photons produced via, $\gamma e^- \rightarrow V' e^-$, with $V'$ decaying to DM which then produces electronic recoil events. In Sec.~\ref{coh} we look at the sensitivity at COHERENT to producing dark photons from neutral pion decay. At COHERENT the rate is dominated by the coherently enhanced nuclear recoil events. In Sec.~\ref{context} we show the derived COHERENT constraints on light DM in the context of the existing constraints on light DM finding that COHERENT excludes previously allowed thermal relic parameter space for $\lesssim 30$ MeV masses. Finally in Sec.~\ref{conc} we conclude and comment on the potential for future limits on the model.

\section{Light DM with Dark Photon Portal}
\label{model}
We assume that the hidden sector $U(1)$ gauge group spontaneously breaks to give the dark photon $V'_{\mu}$ a mass.  Then the relevant terms of the Lagrangian for DM interacting with a kinetically mixed photon are
\be
\mathscr{L} \supset \mathscr{L}_{X} - \frac{1}{4} F'_{\mu \nu}F'^{\mu \nu} + \frac{1}{2} m_{V'}^{2} V'^{\mu} V'_{\mu} - \epsilon F_{\mu \nu} F'^{\mu \nu}
\ee
where the DM portion of the Lagrangian is 
\be
\mathscr{L} = i \bar{X} \slashed{D} X - m_{X} \bar{X} X
\ee
with $D_{\mu} \equiv \partial_{\mu} - i g_{X} V'_{\mu}$ being the covariant derivative and $g_{X}$ the gauge coupling. Strictly speaking we suppose that the dark photon kinetically mixes with the SM hyper-charge field strength, which then induces mixing with both the $Z$ boson and the SM photon after EW symmetry breaking. Throughout, we work directly with the low-energy photon-dark photon mixing parameter $\epsilon$. Details on the procedures for diagonalization and canonical normalization are provided in~\cite{Babu:1997st}.  

In our analysis of TEXONO and COHERENT data we will always assume the mass hierarchy, $ m_{V'} > 2 m_{X}$, such that the decay $V' \rightarrow \bar{X} X$ is allowed. This assumption has ramifications for the thermal relic abundance of DM. First it means that pair annihilation, $\bar{X}X \rightarrow V' V'$, is not permitted. Then the only annihilation mode for DM is $\bar{X}X \rightarrow \bar{f}f$, where $f$ is one of the EM charged particles of the SM.   

The annihilation to EM charged states as predicted in this setup, leads to strong constraints from CMB data~\cite{Finkbeiner:2011dx}. In fact current data is sufficiently strong to completely rule out $s$-wave annihilation for sub-GeV DM if its annihilation is dominated by EM-charged states. Two simple ways out of this conclusion, are to suppress CMB constraints either via assuming (1) $p$-wave annihilating DM, or (2) to introduce a particle/antiparticle asymmetry for DM. While the CMB constraints are completely negligible for $p$-wave annihilation they can still be relevant at limiting the particle-to-antiparticle ratio for asymmetric DM~\cite{Graesser:2011wi,Lin:2011gj,Bell:2014xta}. 

To obtain $p$-wave annihilation the photon portal model would have to invoke scalar DM, while the asymmetric DM case can accommodate either fermionic or scalar DM. For simplicity here we focus on fermionic DM, and expect the derived TEXONO and COHERENT bounds to be similar for scalar DM.

\section{TEXONO's Compton-Like Constraint}
\label{tex}
In this section, we build on the recent work in~\cite{Park:2017prx} which considered dark photon constraints from reactor neutrino experiments. Unlike~\cite{Park:2017prx} however, here we allow for the dark photon to decay to pairs of light DM which then subsequently scatter. 

\begin{figure}[h]
\centering
\includegraphics[height=0.48\textwidth,angle=-90]{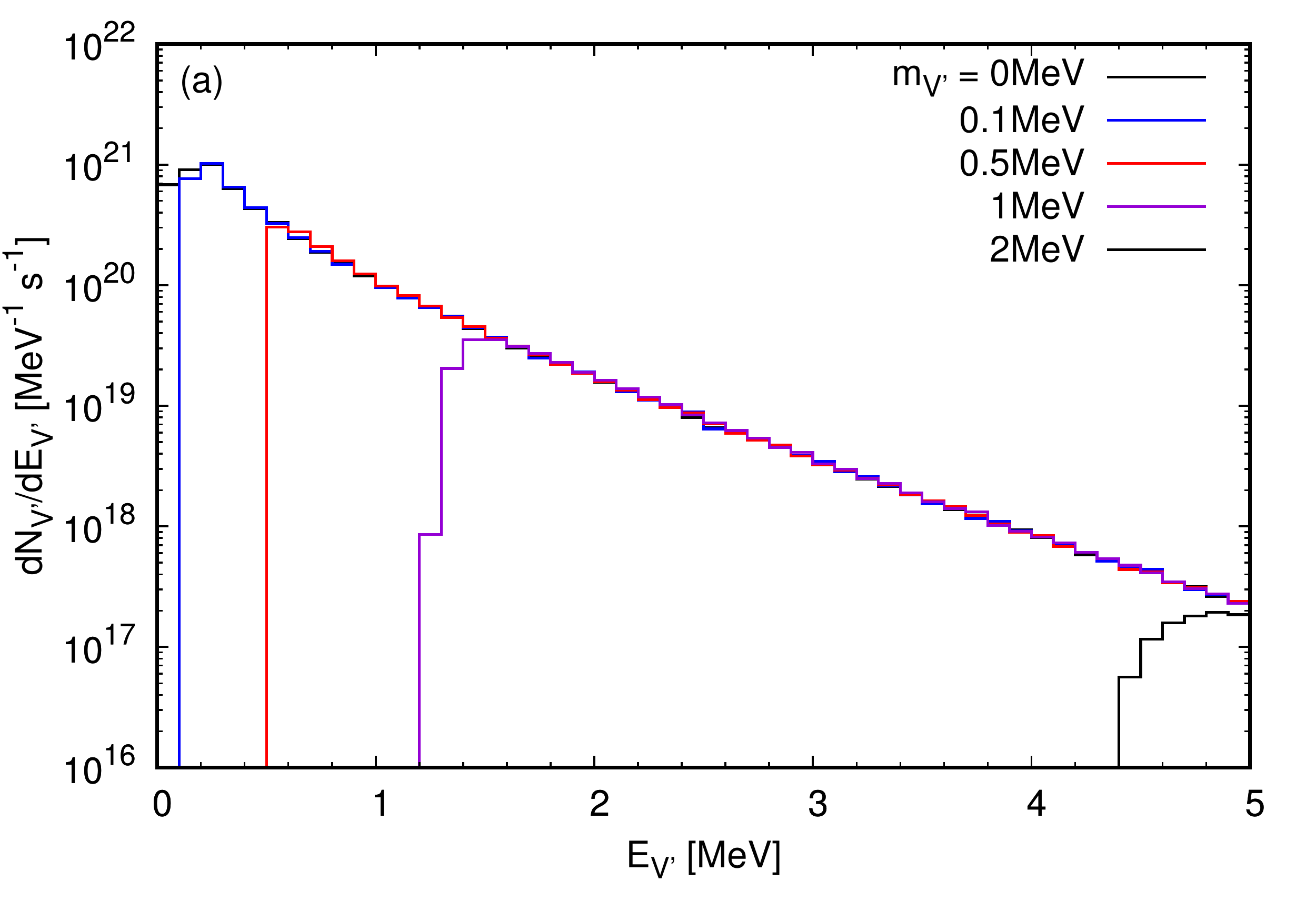}
\includegraphics[height=0.48\textwidth,angle=-90]{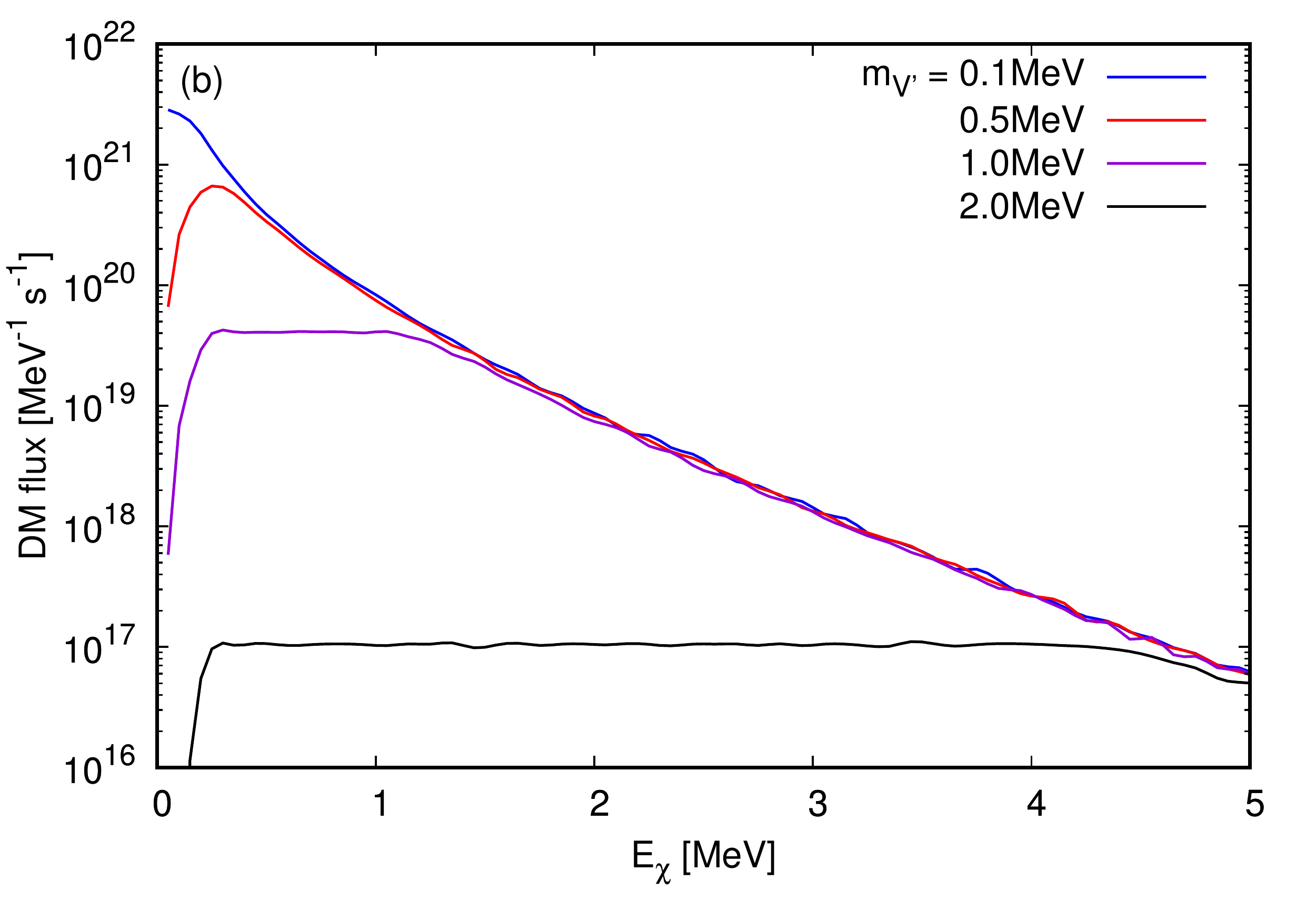}
\includegraphics[height=0.48\textwidth,angle=-90]{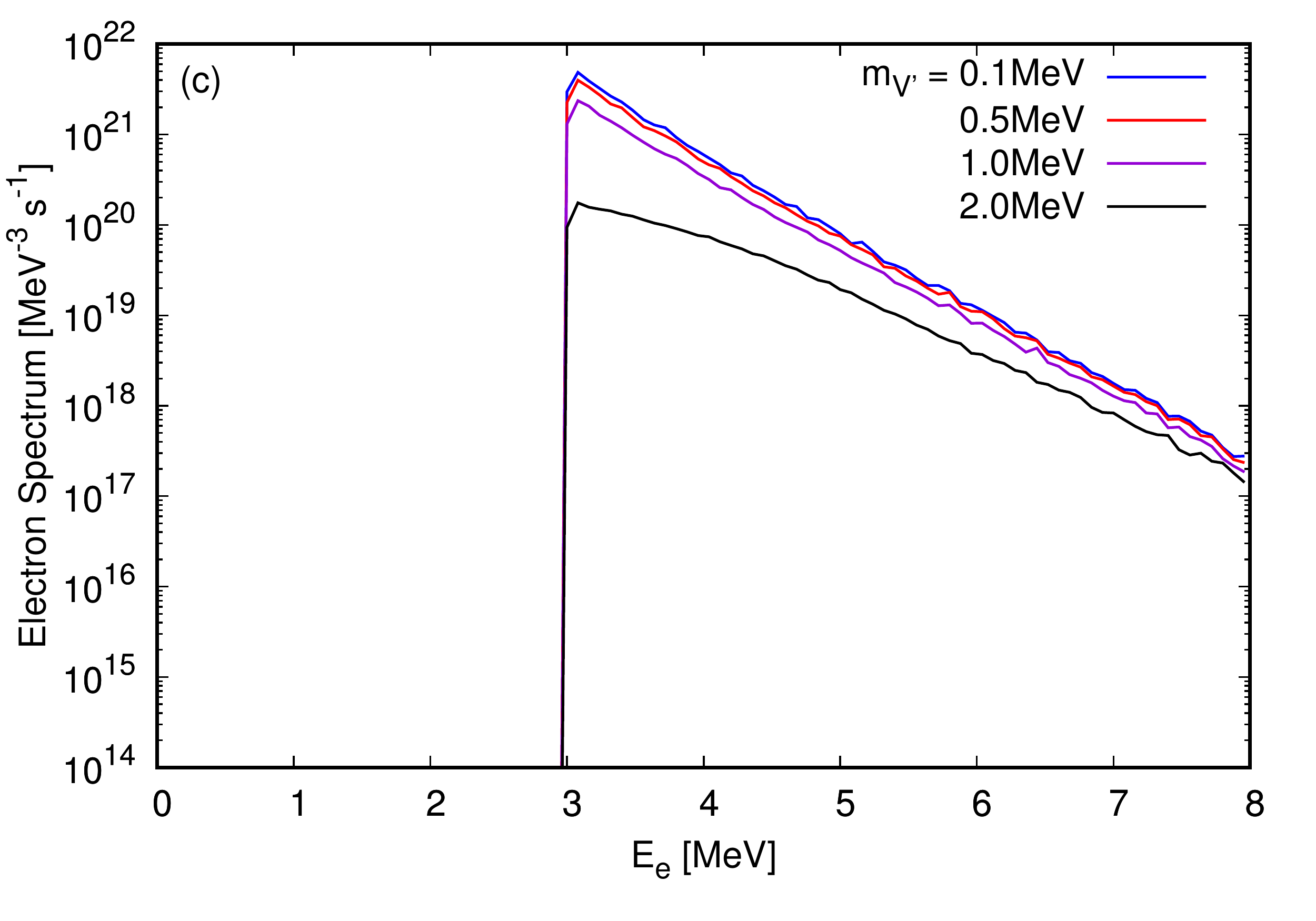}
\includegraphics[height=0.48\textwidth,angle=-90]{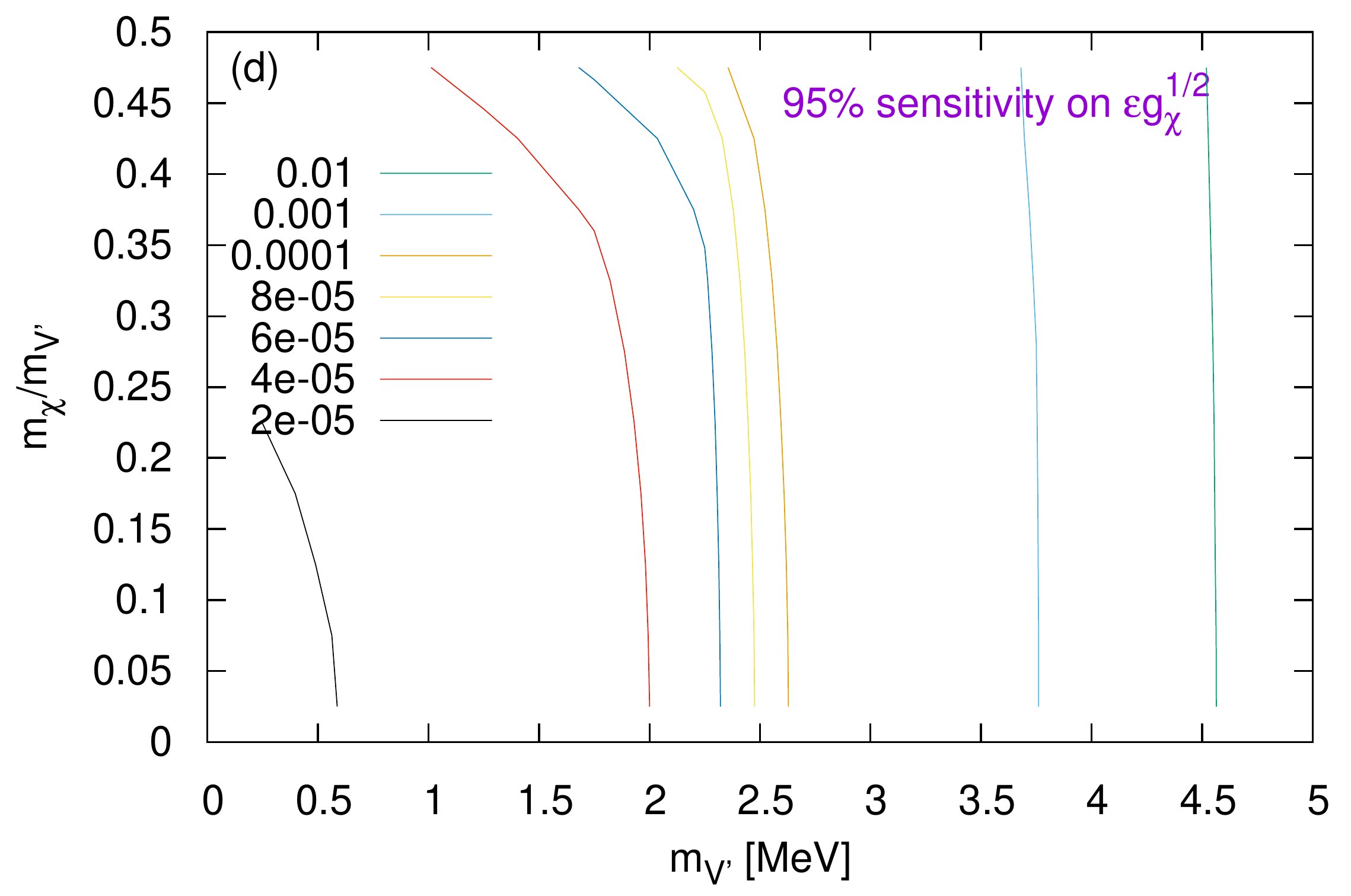}
\caption{The $V'$ flux (a), the DM $\chi$ flux (b), and the detected electron
spectrum (c) from Compton-like scattering
of prompt $\gamma$ on electrons with
$1\,\mbox{GW}$ of thermal reactor power.
For illustration we take $m_{V'} = 0.1, 0.5, 1, 2\,\mbox{MeV}$, $m_\chi = 0$,
the $\gamma$-$V'$ mixing $\epsilon = 1$, and the dark sector gauge coupling $g' = 1$
in (a), (b), and (c). The subplot (d) shows the sensitivity on $\epsilon g_\chi$
for $m_{V'} > 2 m_\chi$.}
\label{fig:Compton}
\end{figure}

The fission process of a thermal reactor can produce a large quantity of prompt~\cite{Park:2017prx}
$\gamma$-rays,
\begin{equation}
  \frac {d N_\gamma}{d E_\gamma}
=
  0.58 \times 10^{21}
  \left( \frac P {\mbox{GW}} \right)
  \exp \left( - \frac {E_\gamma}{0.91\,\mbox{MeV}} \right) \,.
\label{eq:prompt-gamma}
\end{equation}
With a typical thermal power of the order, $P \sim \mathcal O(\mbox{GW})$, around
$\mathcal O(10^{21})$ of photons are produced at $\mathcal O(\mbox{MeV})$ energies.
These $\gamma$'s can scatter with electrons in the reactor to produce
dark photon $V'$ in a Compton-like process, $\gamma e^- \rightarrow V' e^-$.
The dark photon flux is then a convolution of the prompt-$\gamma$ flux
\geqn{eq:prompt-gamma} and the differential cross section
$d \sigma_{\gamma e^- \rightarrow V' e^-}/d E_{V'}$,
\begin{equation}
  \frac {d N_{V'}}{d E_{V'}}
=
  \int \frac {d N_\gamma}{d E_\gamma}
       \frac {d \sigma_{\gamma e^- \rightarrow V' e^-} (E_\gamma)}{\sigma_{tot} d E_{V'}}
  d E_\gamma \,.
\end{equation}
Note that the differential cross section is normalized by the total cross section
$\sigma_{tot}$ of Compton process which dominates the interaction of prompt
$\gamma$-rays inside the reactor.
In \gfig{fig:Compton}(a) we show the $V'$ flux
for $1\,\mbox{GW}$ of thermal reactor power and $\gamma$-$V'$ mixing
$\epsilon = 1$. The dark photon mass sets a natural cut on its flux.
From massless $V'$ to $m_{V'} = 1\,\mbox{MeV}$, the flux
drops by almost two orders. Note that the cross section
$\sigma_{\gamma e^- \rightarrow V' e^-}$ of Compton-like process
is proportional to $\epsilon^2$.

The dark photon $V'$ then decays into a pair of DM particles if
$m_{V'} > 2 m_\chi$. Since the $V'$-$\chi$ coupling is much larger
than the $V'$-$e$ coupling, the dark photon $V'$ mainly decays into
a pair of $\chi$. The $\chi$ flux is then a further convolution,
\begin{equation}
  \frac {d N_\chi}{d E_\chi}
=
  \int
  \frac {d N_{V'}}{d E_{V'}}
  \frac {d \Gamma_{V' \rightarrow \chi \chi} (E_{V'})}{\Gamma d E_\chi}
  d E_{V'} \,.
\label{eq:Nchi}
\end{equation}
Since the $V'$ decay is isotropic in its rest frame, the DM energy $E_\chi$
is evenly distributed within the kinematically allowed range
\begin{equation}
  E_- \leq E_\chi \leq E_+ \,,
\qquad \mbox{where} \qquad
  E_\pm
=
  \frac 1 2 E_{V'}
\pm
  \frac 1 2 \sqrt{E^2_{V'} - m^2_{V'}} \sqrt{1 - \frac {4 m^2_\chi}{m^2_{V'}}} \,.
\end{equation}
Consequently, the flux cut is not present in the \gfig{fig:Compton}(b)
of DM flux and the DM energy $E_\chi$ can stretch to much lower scale,
especially when $m_\chi \ll m_{V'} \ll E_{V'}$.
Since the $V'$ decay width is normalized as $d \Gamma / \Gamma$, the DM flux
$d N_\chi / d E_\chi$ is also proportional to $\epsilon^2$, the same
as $d N_{V'} / d E_{V'}$.

The scattered electron spectrum at the TEXONO experiment
\begin{equation}
  \frac {d N_e}{d E_e}
=
  \frac {N_e T}{4 \pi R^2}
  \int \frac {2 d N_\chi}{d E_\chi}
       \frac {d \sigma_{\chi e^- \rightarrow \chi e^-} (E_\chi)}{d E_e}
  d E_\chi
\label{eq:Ne}
\end{equation}
does not
drop that much when $m_{V'}$ increases from $0\,\mbox{MeV}$ to $1\,\mbox{MeV}$.
This is because the TEXONO detection threshold is even higher. Only events with
electron energy larger than $3\,\mbox{MeV}$ can be recorded. As shown in
\gfig{fig:Compton}(c), the detection electron spectrum
is not significant affected for $m_{V'} \lesssim 1\,\mbox{MeV}$. The decrease becomes
visible when $m_{V'}$ increases to $2\,\mbox{MeV}$. Note that the scattering
cross section has $\epsilon^2 g^2_\chi$ dependence. In total, the scattered
electron spectrum $d N_e / d E_e$ is scaled by a factor of $\epsilon^4 g^2_\chi$.
The factor of $2$ associated with $d N_\chi / d E_\chi$ comes from the fact that
a single $V'$ can produce two DM particles, $\chi$ and $\bar \chi$. Both of them
can scatter with electron via $t$-channel $V'$ mediation.

The prefactors in \geqn{eq:Ne} takes into account the number of electrons $N_e$,
the integrated run time $T$, and the dilution factor $1/4 \pi R^2$ of DM flux due
to Gauss law which is a valid approach as long as the decay length of $V'$
is negligibly small.
For the TEXONO experiment, $187\,\mbox{kg}$ CsI(Tl) scintillating
crystal detector is placed 28 meters from the core of a $2.9\,\mbox{GW}$
thermal-power reactor. In the energy range $3\,\mbox{MeV} \leq E_e \leq 8\,\mbox{MeV}$
of recoiled electrons, the TEXONO has collected $414 \pm 80 (stat) \pm 61 (sys)$ events
\cite{Deniz:2009mu}. We fit the beam-on data in the Fig.16(b) of \cite{Deniz:2009mu}
with both SM and dark matter contributions
\begin{equation}
  \chi^2
\equiv
  \sum_i \left( \frac {f_{SM} N^{SM}_i + f_\chi N^\chi_i - N^i_{exp}}{\Delta_i} \right)^2
+ \left( \frac {f_{SM} - 1}{\Delta_{SM}} \right)^2 \,,
\label{eq:chi2}
\end{equation}
where $N^{SM}_i$ is the SM event number inside
the $i$-th bin while the DM counterpart
$N^\chi_i$, and $N^{exp}_i$ is the observed event
number in the $i$-th bin, and $\Delta_i$ is the corresponding uncertainty.
On the other hand, $f_{SM}$ is the normalization for the SM contribution and
$f_\chi$ is the DM counterpart.
Since the DM contribution always scales with $\epsilon^4 g^2_\chi$,
we define $f_\chi \equiv \epsilon^4 g^2_\chi$ and
$N^\chi_i \equiv N^\chi_i (m_{V'}, m_\chi, \epsilon g^{1/2}_\chi = 1)$
for convenience.
If the SM is complete description, the normalization factors reduce to
$f_{SM} = 1$ and $f_\chi = 0$.
Both $N^{exp}_i$ and $\Delta_i$ are read off from the beam data in the Fig.16(b) of
\cite{Deniz:2009mu}. To account for the systematics we also add
a nuisance term with $\Delta_{SM}$. We assign $\Delta_{SM} = 5\%$ which
is quite conservative since the SM prediction for neutrino elastic scattering with
electron is very precise.
During a single fit, we fix the four DM sector parameters
$(m_{V'}, m_\chi)$ and adjust the two
normalizations $f_{SM}$ and $f_\chi$. This would lead to a two dimensional
contour in the $f_{SM}$--$f_\chi$ plan.
Using analytic $\chi^2$ fit technique \cite{Ge:2016zro}, the marginalized
uncertainty $\Delta f_\chi$ of $f_\chi$, or upper limit
$f_\chi \leq \Delta f_\chi$ at certain confidence level, can then be expressed in terms
of sensitivity on $\epsilon g^{1/2}_\chi$,
\begin{equation}
  \epsilon g^{1/2}_\chi
\leq
  (\Delta f_\chi)^{\frac 1 4} \,.
\end{equation}

We show the sensitivity the $95\%$ limit on the couplings $\epsilon g^{1/2}_\chi$
in \gfig{fig:Compton}(d). The dependence of sensitivities on $m_{V'}$ is minor for
$m_{V'} \lesssim 2\,\mbox{MeV}$ and decreases much faster when $m_{V'}$ goes
beyond $2\,\mbox{MeV}$. Given $m_{V'}$, the sensitivity is almost independent
of the DM mass $m_\chi$, especially for $m_\chi < m_{V'} / 4$ when
$m_{V'} \lesssim 2\,\mbox{MeV}$ and $m_\chi < 0.4 \times m_{V'}$ when
$m_{V'} > 2\,\mbox{MeV}$.
It is interesting to see that the TEXONO measurement of electron recoil can already
constrain the dark coupling to $\epsilon g_\chi$ as small as $\mathcal O(10^{-5})$
for $m_{V'} \approx 1\,\mbox{MeV}$.

\section{COHERENT Constraint}
\label{coh}

The COHERENT experiment recently measured for the first time the neutrino-nucleus coherent scattering
 \cite{Akimov:2017ade} with neutrinos produced by a source of stopped charged pions.
It utilizes $1.76 \times 10^{23}$ of
$\sim 1\,\mbox{GeV}$ protons striking a mercury target.
As a byproduct, the same proton beam can also produce neutral pions which mainly
decays into a pair of photons. We use the analytic
expressions in \cite{Burman89} to estimate the spectrum of neutral pions,
$d N_{\pi^0} / d E_{\pi^0}$, and the normalization is fixed by the experimental
estimation of $~0.08$ $\pi^+$ for each proton according to the COHERENT experiment
\cite{Akimov:2017ade}.

\begin{figure}[t]
\centering
\includegraphics[height=0.48\textwidth,angle=-90]{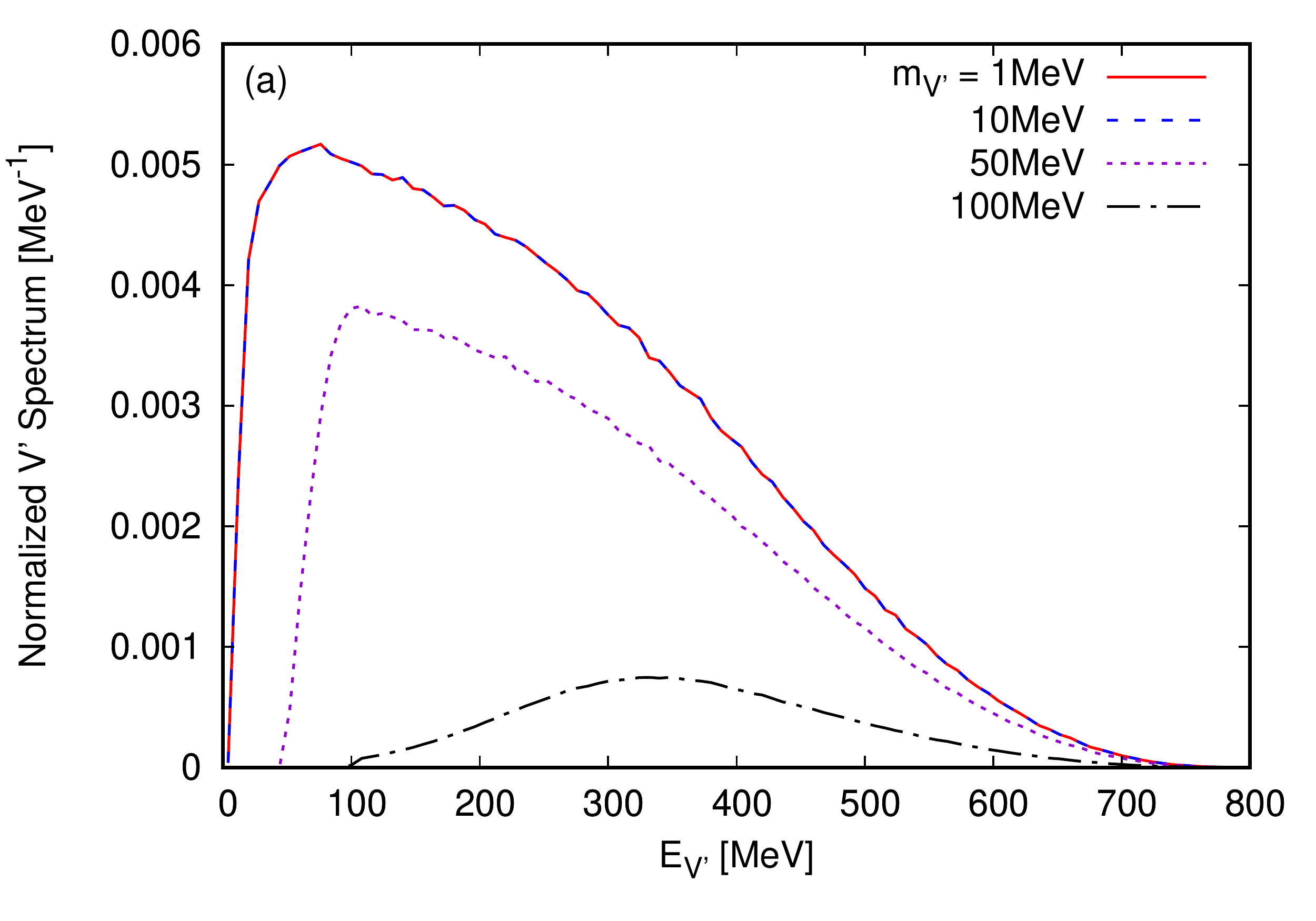}
\includegraphics[height=0.48\textwidth,angle=-90]{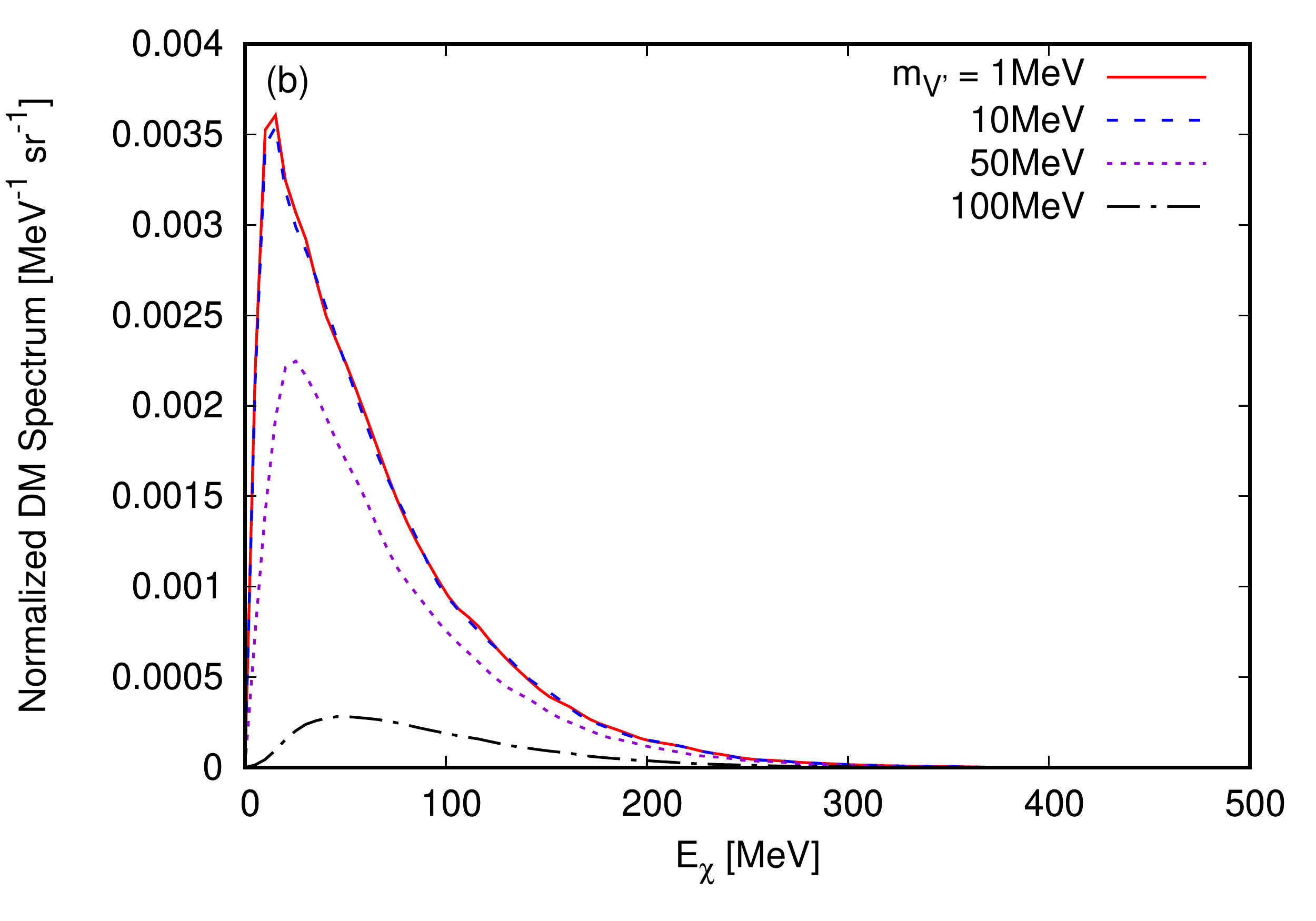}
\includegraphics[height=0.48\textwidth,angle=-90]{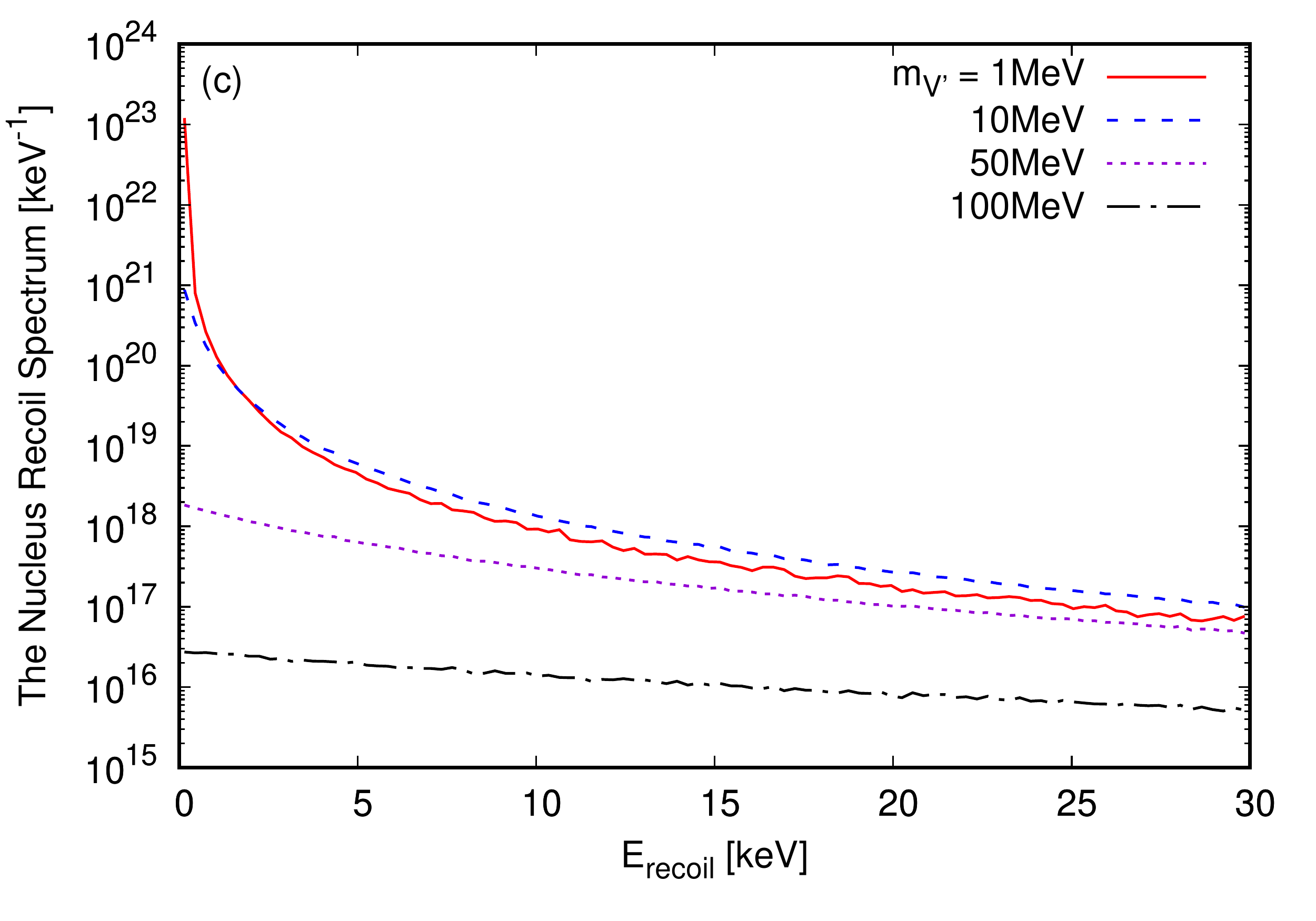}
\includegraphics[height=0.48\textwidth,angle=-90]{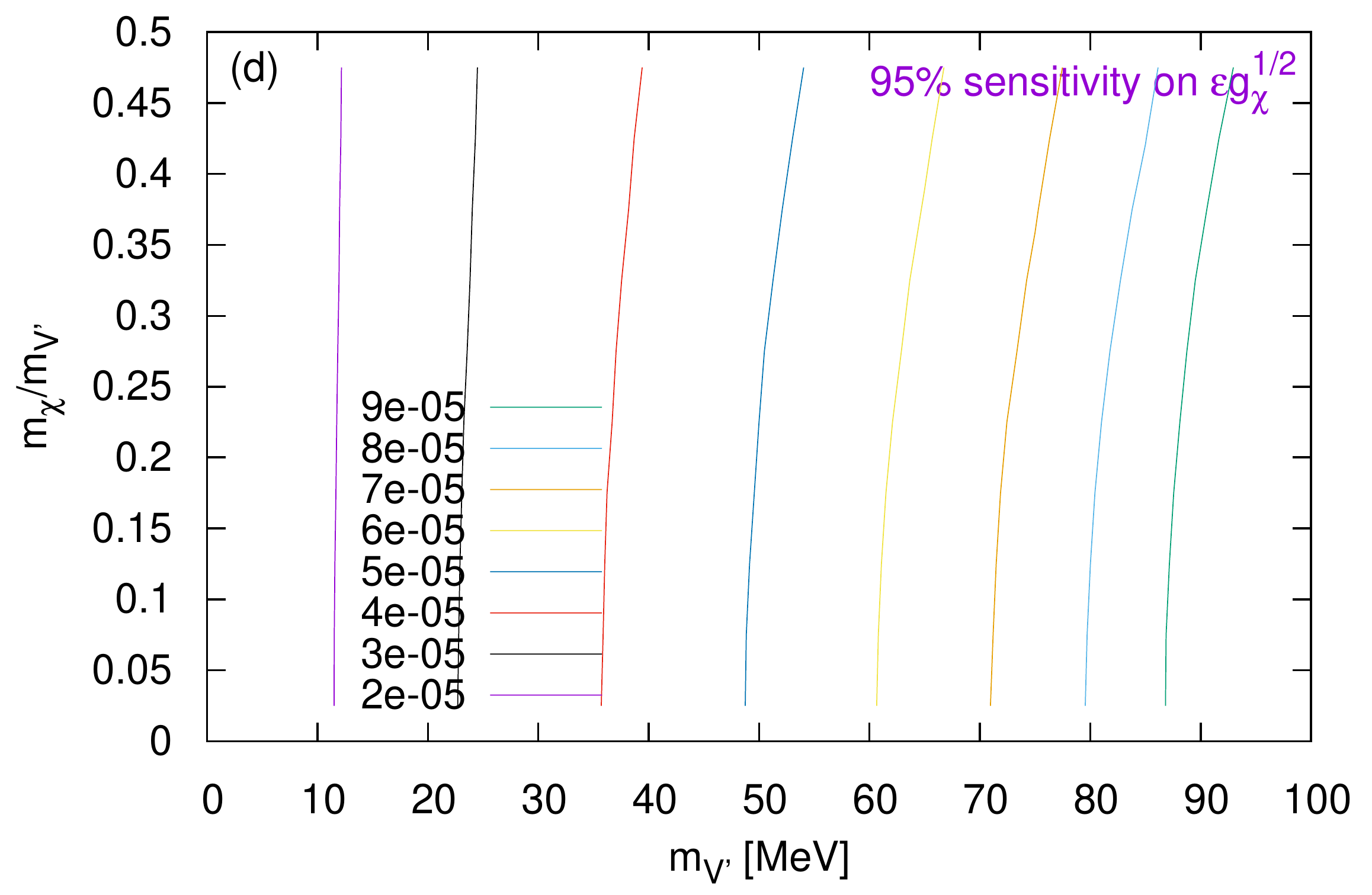}
\caption{The normalized fluxes for $\chi$ (a) from $\pi^0$ decay at the COHERENT experiment. On the detection side, (b) shows the recoil energy spectrum in 308 live-days of data collection with 14.6\,kg of CsI[Na] while (c) and (d) shows the corresponding 95\% sensitivity on the DM coupling
$\epsilon g^{1/2}_\chi$. In the first two subplots, the DM mass is
simply set to zero, $m_\chi = 0$, and $\epsilon = 1$ as well as $g' = 1$.}
\label{fig:fixedTarget}
\end{figure}

With $\gamma$-$V'$ mixing, neutral pions have a
non-negligible probability of decay into the dark photon, $\pi^0 \rightarrow \gamma V'$. 
The dark photon spectrum is then a convolution
of the $\pi^0$ spectrum and the differential branching ratio of the
$\pi^0 \rightarrow \gamma V'$ decay,
\begin{equation}
  \frac {d N_{V'}}{d E_{V'} d \cos \theta_{V'}}
=
  \int \frac {d N_{\pi^0}}{d E_{\pi^0} d \cos \theta_{\pi^0}}
       \frac {d \Gamma_{\pi^0 \rightarrow \gamma V'} (E_{\pi^0})}
             {\Gamma_{\pi^0} d E_{V'} d \cos \bar \theta_{V'} d \bar \phi_{V'}}
  d E_{\pi^0} d \cos \theta_{\pi^0} d \phi_{V'} \,,
\end{equation}
where $\bar \theta_{V'}$ and $\bar \phi_{V'}$ are the $V'$ zenith and azimuthal angles
in the $\pi^0$ frame while $\theta_{V'}$ and $\phi_{V'}$ are the counterparts in the
lab frame.
In \gfig{fig:fixedTarget}(b) we show the normalized $V'$ spectrum with different mediator masses,
$m_{V'} = 1$, $10$, $50$, $100\,\mbox{MeV}$. Although there is a kinematic cut
$E_{V'} > m_{V'}$, the spectrum goes to zero smoothly. From $m_{V'} = 1\,\mbox{MeV}$
to $m_{V'} = 100\,\mbox{MeV}$, the spectrum peak drops by a factor of $5 \sim 6$.
The dark photon $V'$ further decays into a pair of dark matter particles.
\begin{equation}
  \frac {d N_\chi}{d E_\chi d \cos \theta_\chi}
=
  \int \frac {d N_{V'}}{d E_{V'} d \cos \theta_{V'}}
       \frac {d \Gamma_{V' \rightarrow \chi \chi} (E_{V'})}
             {\Gamma_{V'} d E_\chi d \cos \bar \theta_\chi d \bar \phi_\chi}
  d E_{V'} d \cos \theta_{V'} d \phi_\chi \,.
\end{equation}
Similarly, $\bar \theta_\chi$ and $\bar \phi_\chi$ are the $\chi$ zenith
and azimuthal angles in the $V'$ frame while $\theta_\chi$ and $\phi_\chi$
are the counterparts in the lab frame. In this way, we take into account the
fact that the COHERENT system is not isotropic. Then we fix the off-axis angle
$\theta_\chi = 110^\circ$ for the DM flux hitting on the target.
We show the normalized DM spectrum in \gfig{fig:fixedTarget}(c).

Here the DM
$\chi$ from $V'$ decay scatters with nuclei via $V'$ mediation, in a manner analogous to the scattering with electrons in the TEXONO experiment considered in Sec.~\ref{tex}. 
In the coherent region, the cross section,
\begin{equation}
  \frac {d \sigma}{d E_r}
=
  \frac {Q^2_{eff}(q^2) F^2(q^2)}{8 \pi (E^2_\chi - m^2_\chi)}
\left[
  2 (g^2_V + g^2_A) M_N E^2_\chi \left( 1 - \frac {E_r}{E_\chi} - \frac {M_N E_r}{2 E^2_\chi} \right)
- g^2_A m^2_\chi (E_r + 2 M_N)
\right] \,,
\end{equation}
can feel the effective charge of the nucleus collectively,
$Q_{eff} = \epsilon Z e / (q^2 - m^2_{V'})$.
Consequently, the cross section is modulated by $Z^2$ instead of $Z$ and hence is
significantly enhanced.
The recoil energy $E_r \equiv q^2 / 2 M_N$ is proportional to the momentum transfer
$q^2$, suppressed by the nucleus mass $M_N$, and constrained within the range
\begin{equation}
  0
\leq
  E_r
\leq
  \frac {2 M_N (E^2_\chi - m^2_\chi)}{m^2_\chi + M^2_N + 2 M_N E_\chi}
\approx
  2 \frac {E^2_\chi - m^2_\chi}{M_N} \,.
\end{equation}
Since the nucleus mass $M_N \sim \mathcal O(100\mbox{GeV})$
is much larger than the DM mass $m_\chi \sim \mathcal O(100\mbox{MeV})$, the recoil energy is significantly
suppressed to $\mbox{keV}$ energies as shown in \gfig{fig:fixedTarget}\,(c).
Since the COHERENT experiment
uses 14.6\,kg CsI target for detection, such that the contribution of coherent scattering on both Cs and I must be included. 

The recoil energy spectrum
can be directly confronted with the COHERENT data
to give constraints on the DM couplings. On top of the SM
neutrino background that is shown as the shaded histogram in
Fig.3 of \cite{Akimov:2017ade} we add the DM contribution to
fit the COHERENT data points in the same figure.
In \gfig{fig:fixedTarget}\,(d) we show the 95\% sensitivity constraint on the DM coupling $\epsilon g^{1/2}_\chi$. We use the $\chi^2$ definition in \geqn{eq:chi2}, but with $\Delta_{SM} = 10\%$ \cite{Coloma:2017egw}. Interestingly, the
result is almost independent of the DM mass $m_\chi$ for a
given dark photon mass $m_{V'}$. For $m_{V'} \lesssim 40\,\mbox{MeV}$, the curves are nearly vertical. In other
words, the sensitivity does not change much when varying 
$m_\chi / m_{V'}$ or equivalently $m_\chi$. Only
for mediators masses $m_{V'} \gtrsim 50\,\mbox{MeV}$, does the sensitivity show
some slight dependence on $m_\chi$.

\section{Comparison with other constraints: COHERENT at the Direct Detection Frontier}
\label{context}
A number of other earlier experiments also constrain the same parameter space for photon portal DM. Some of these include LSND, BaBar, and XENON10 data. 

For illustration we map these bounds into the $\bar{\sigma}_{e}$-$m_{X}$ plane where $\bar{\sigma}_{e}$ represents the DM-electron cross section. This cross section is 
\begin{equation}
\bar{\sigma}_{e} =
    \begin{cases*}
     16 \pi \alpha \alpha_{X} \epsilon^{2} \mu_{eX}^{2}/m_{V'}^{4} & ($m_{V'} \gg m_{X} v$) \\
        16 \pi \alpha \alpha_{X} \epsilon^{2}\mu_{eX}^{2} / (m_{e}\alpha)^{4}     &  ($m_{V'} \ll m_{X} v$),
            \end{cases*}
  \end{equation}
where in the latter case the atomic electron momentum dominates the momentum exchange, and $\mu_{eX}$ is the electron-DM reduced mass. In the mass range of interest, we are always in the regime where $\bar{\sigma}_{e} \propto m_{V'}^{-4}$. 

\begin{figure*}[t!]
\centering
\includegraphics[height=0.48\textwidth]{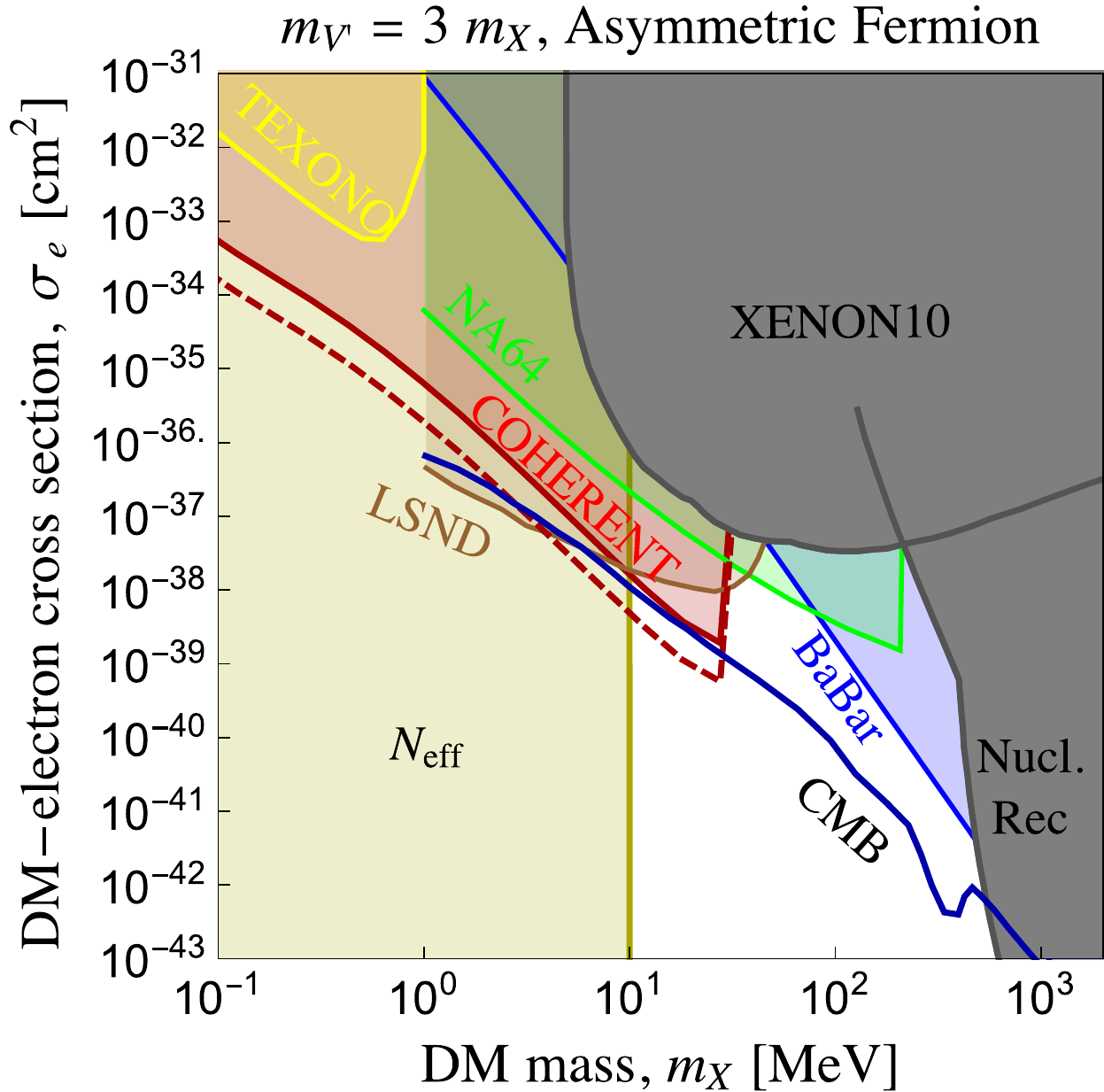}
\includegraphics[height=0.48\textwidth]{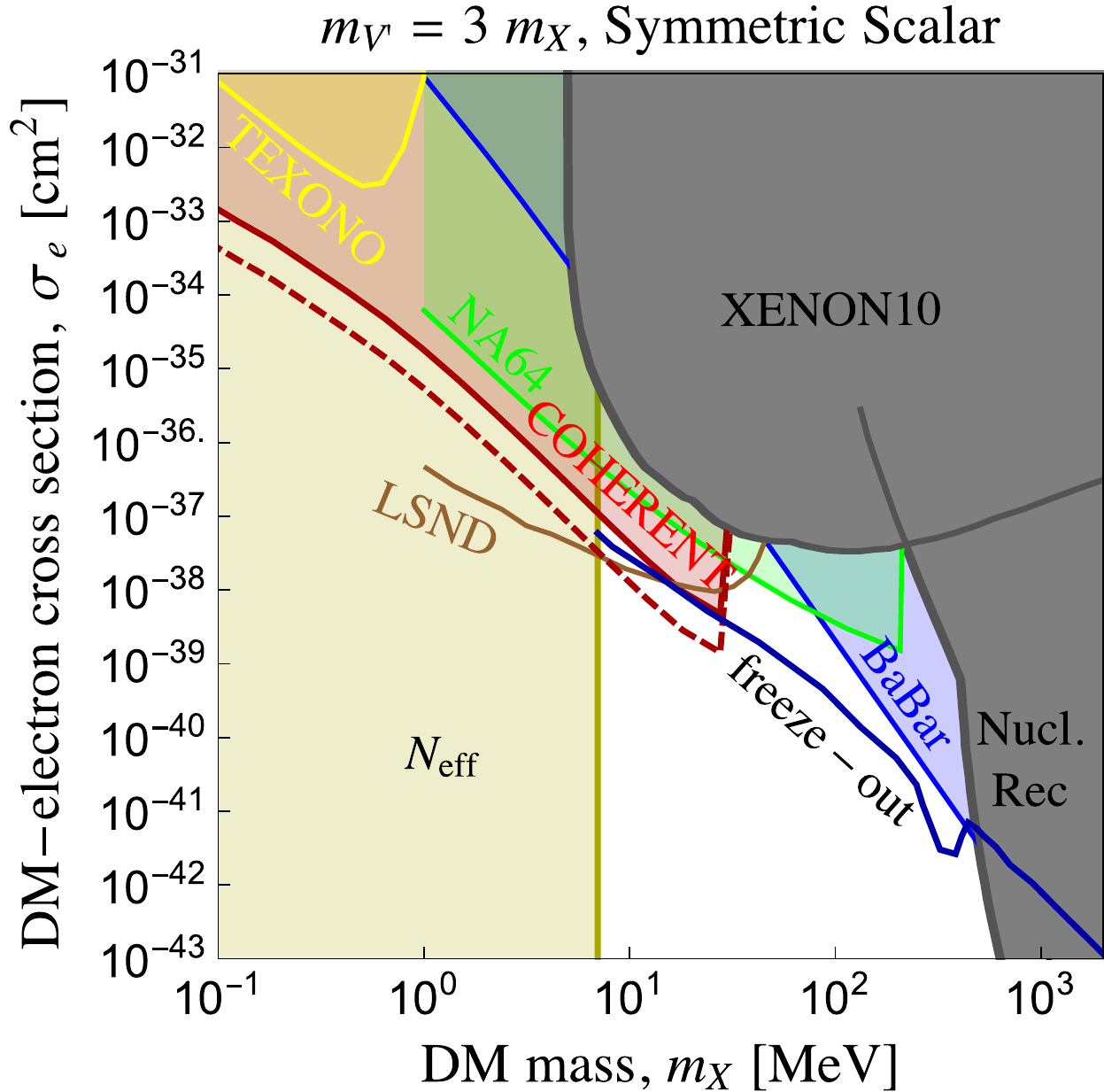}
\caption{The COHERENT bounds derived in this work in the context of other bounds on DM interacting with a kinetically mixed dark photon.
The left and right panels standards for asymmetric fermion and symmetric scalar dark matters, respectively. The red solid curve show the COHERENT bound with $g_{X}$ at the cluster bounds on DM self-interactions, and dashed red with $g_{X}$ at 0.3 times the cluster bound. One can re-scale the BaBar and NA64 bounds for other choices of the dark coupling straightforwardly since the cross section $\propto g_{X}^{2}$.}
\label{fig:cross}
\end{figure*}

To compare with existing constraints, we map our bounds on the dark photon couplings to the $\bar{\sigma}_{e}$ plane in Fig.~\ref{fig:cross}. The other displayed constraints include the direct XENON10 bounds~\cite{Essig:2012yx}, BaBar's $e^{+}e^{-} \rightarrow \gamma + {\rm invisible}$ search~\cite{Aubert:2008as,Essig:2013lka,Essig:2013vha,Lees:2017lec}, LSND~\cite{Essig:2015cda} (see also \cite{Kahn:2014sra,deNiverville:2011it,Batell:2009di}), and $N_{{\rm eff}}$ bounds on the CMB and BBN era radiation density~\cite{Boehm:2013jpa}. In addition the experiment NA64 searches for invisible decays of the dark photon and recently established updated limits~\cite{Banerjee:2017hhz}. We note as well that the recent updates to low-mass DM-nuclear scattering are quite strong and can also be mapped into the above plane~\cite{Angloher:2015ewa,Kouvaris:2016afs,McCabe:2017rln}. Lastly, as in Ref.~\cite{Essig:2015cda} we require that the DM coupling $g_{\chi}$ not violate bounds on the DM self-scattering cross section. In particular we impose the cluster bound on DM self-interactions~\cite{Randall:2007ph} $\sigma_{DM}/m_{X} \lesssim 1~{\rm cm}^{2}/{\rm g}$ throughout, and ensure that each curve in Fig.~\ref{fig:cross} obeys this bound when a given experimental curve does not directly constrain the dark coupling itself. Notice that the $\bar{\sigma}_{e} \propto g_{\chi}^{2} \epsilon^{2}$, whereas COHERENT bounds the combination $\epsilon^{2} g_{\chi}$. Therefore to draw the COHERENT bound in the $\bar{\sigma}_{e}$ plane one must make an explicit choice for $g_{\chi}$. We take $g_{\chi}=1$ in Fig.~\ref{fig:cross} to show conservative constraints as solid lines and $g_\chi$ at 0.3 times the cluster bound to show the moderate enhancement as dashed lines.

As we can see from Fig.~\ref{fig:cross}, the COHERENT data is already sufficiently strong to probe thermal relic DM at a level not probed by other experiments. In particular for $m_{V'} = 3 m_{X}$, COHERENT excludes thermal relic cross sections for $m_{X} \lesssim 30$ MeV. Since the BaBar constraint constrains thermal relics to have masses, $m_{X}<800$ MeV, there remains a window of allowed masses $30\lesssim m_{X} \lesssim 800$ MeV for this class of models to successfully account for the relic abundance via the standard freeze-out mechanism. Future COHERENT data as well as future direct detection limits~\cite{Essig:2015cda,Hochberg:2015pha,Hochberg:2015fth,Hochberg:2016ntt,Derenzo:2016fse,Essig:2017kqs,Tiffenberg:2017aac,Mei:2017etc} on electron scattering may probe the remaining range of allowed masses.

For comparison, we also derive the bounds for symmetric scalar DM and show
the results in the right panel of \gfig{fig:cross}.
The COHERENT constraint on the symmetric scalar DM is similar to the one
on the asymmetric fermionic DM. Comparing with other experiments and observations,
the COHERENT constraint is still much better around $(10 \sim 30)$\,MeV.

\section{Conclusion}
\label{conc}

In this paper we studied the implications of recent data from both TEXONO and COHERENT for the photon portal interaction of light DM. In TEXONO's case the dark photon is produced in a Compton-like process, $\gamma e^- \rightarrow V' e^-$, and $V'$ subsequently decays to pairs of DM particles which then scatter off the electrons. For COHERENT the neutral pion decays provide the dark photon production, $\pi^{0} \rightarrow \gamma + V'$, and the coherent enhanced nuclear scattering provides the main detection method. 

In the future we expect COHERENT's sensitivity to improve considerably. In addition to the raw exposure, the collaboration plans to use additional nuclear targets beyond the CsI used in the present dataset. For example, if anomalous DM induced events were present in future data, the correlation with the predicted $\sim Z^{2}$ coherent enhancement with different nuclei would provide an additional handle for discriminating the presence of DM. In addition, imposing a timing cut can reduce the background since the DM signal comes from neutral pion decay but the background is from charged pions/muons with very different lifetimes \cite{Coloma:2017egw}. The timing cut can further improve the COHERENT sensitivity.

We also note that the reactor based coherent neutrino scattering experiment MINER~\cite{Agnolet:2016zir} will soon take data as well. Although the reactor power at MINER will be orders of magnitude below  TEXONO's, the significant reduction in energy threshold expected by MINER may open up new parameter space for them, though only at sub-MeV DM masses.

\noindent {\bf \Large Acknowledgements} 

SFG would like to thank Juan I. Collar for providing the COHERENT data points and H.K. Park for email correspondences. SFG is grateful to Hong-Jian He for his support and hospitality of visiting the T.D. Lee Institute (TDLI)
at the Shanghai Jiao Tong University, where this work was presented at the {\it International Workshop on WIMP Dark Matter and Beyond} on September 18, 2017 (\href{https://users.hepforge.org/~gesf/talks/170918_LeeInstitute_lightDM.pdf}{link}),
and the Center for High Energy Physics at Tsinghua University where part of this
paper was prepared.
This work was supported by World Premier International Research Center Initiative (WPI Initiative), MEXT, Japan.
SFG was supported by JSPS KAKENHI Grant Number JP18K13536.
IMS would like to thank the USD Physics Department for its support. We are also grateful to Rouven Essig and Sergei Gninenko for providing the constraints on BaBar and NA64 respectively.

\bibliographystyle{JHEP}

\bibliography{nuExpDM}

\end{document}